# Propensity-score matching analysis in COVID-19-related studies: a method and quality systematic review

Chunhui Gu[a,b], Email: Chunhui.gu@uth.tmc.edu
Ruosha Li [b], Email: ruosha.li@uth.tmc.edu
Guoqiang Zhang [a, *]
[a] Department of Neurology, McGovern Medical School, University of Texas Health Science Center at Houston, Texas, United States
[b] Department of Biostatistics and Data Science, School of Public Health, University of Texas Health Science Center at Houston, Texas, United States

\*  Corresponding author:
The University of Texas Health Science Center at Houston
1133 Freeman Blvd | Houston, TX 77030
Tel: 713-500-7117
Email: guo-qiang.zhang@uth.tmc.edu

# Abstract

**Objectives:** To provide an overall quality assessment of the methods used for COVID-19-related studies using propensity score matching (PSM).

**Study Design and Setting:** A systematic search was conducted in June 2021 on PubMed to identify COVID-19-related studies that use the PSM analysis between 2020 and 2021. Key information about study design and PSM analysis were extracted, such as covariates, matching algorithm, and reporting of estimated treatment effect type.

**Results:** One-hundred-and-fifty (87.72%) cohort studies and thirteen (7.60%) case-control studies were found among 171 identified articles. Forty-five studies (26.32%) provided a reasonable justification for covariates selection. One-hundred-and-three (60.23%) and Sixty-nine (40.35%) studies did not provide the model that was used for calculating the propensity score or did not report the matching algorithm, respectively. Seventy-three (42.69%) studies reported the method(s) for checking covariates balance. Forty studies (23.39%) had a statistician co-author. All the case-control studies (n=13) did not have a statistician co-author (p=0.006) and all studies that clarified the treatment effect estimation (n=6) had a statistician co-author (p<0.001).

**Conclusions:** The reporting quality of the PSM analysis is suboptimal in some COVID-19 epidemiological studies. Some pitfalls may undermine study findings that involve PSM analysis, such as a mismatch between PSM analysis and study design.

Abstract word count: 200

Running title: Evaluating propensity-score matching analysis in COVID-19-related studies

> What is new?
>
> Key findings
> - A non-neglectable number of Covid-19 studies that use propensity score matching (PSM) analysis are case-control studies, while the appropriateness of the application of such a method is questionable.
> - There is a lack of reporting estimated treatment effect type from the PSM analysis used in a study and this may be a result of the lack of understanding of the different types of treatment effects.
>
> What this adds to what is known?
> - The average treatment effect in treated (ATT) that arrived from the nearest neighbor matching (NNM), which is the most dominant matching algorithm, may not be optimal in serving the purpose of a typical Covid-19 studies.
>
> What is the implication and what should change now?
> - In addition to a continuous effect in improving the reporting quality of the propensity score analysis, attention should also be given to whether the implementation of the PSM analysis appropriately fits a study design and the purpose of a study.

## Introduction

In medicine, it is common that an effect of a factor is of interest and such an effect can be therapeutic, preventive, or detrimental. Such factors are called treatments, exposures, or interventions depending on the context and purpose when it is offered. In analytical models, we often use these terms interchangeably. Usually, along with the treated group, a control group, target population, and outcome(s) of interest are also specified to form well-designed comparative effectiveness research (CER) [1] under the PICO framework [2]. Well-implemented Randomized Clinical Trials (RCTs) remain the gold standard for this type of study [3]. When perfectly implemented, subjects are randomly allocated into different treatment groups and balance is achieved both in measured baseline characteristics and unmeasured baseline characteristics. Therefore, observed differences in outcomes between different treatment groups can be exclusively attributed to the difference between treatments. However, RCTs can be lengthy, resource-intensive, and may involve challenging ethical issues [4] [5].

Instead, observational studies are commonly used as a first-line substitution in exploring a potential effect of a treatment. RCTs can be introduced subsequently if relatively strong evidence is observed. Unlike in RCTs, subjects in observational studies are not assigned to different treatments at random [6]. The treatment assignment is usually associated with baseline characteristics and confounders are introduced inevitably. The propensity score (PS) analysis is a popular method for adjusting confounders in observational studies [7]. A propensity score, which is defined as the probability of receiving a treatment based on a subject's measured baseline characteristics [8], will be estimated for each subject in a study. The estimation is through a model that uses measured baseline characteristics (called covariates in the PSM analysis) as predictors and treatment assignment as the outcome. Such a model is called the propensity score model to differentiate it from the analysis model that estimates the effect of treatment on the outcome of interest.

There are different ways to use the propensity score including matching, weighting, stratification, and adjustment[9]. The propensity score matching (PSM) analysis is the most popular method for its similarity to RCTs in design and analysis [10]. After matching properly based on the propensity score, the measured baseline covariates will be balanced between the treatment group and the control group. Theoretically, when covariates used in the propensity score model included all the confounders, the confounders in data after matched should be balanced and the rest of the analysis can be done similarly as it is in the RCTs [9]. Being relatively simple to use and easy to interpret the result makes PSM analysis popular in the medical field.

The emergence of electronic health record (EHR) databases and EHR-linked registries makes it much easier to conduct observational studies than ever before. More information can potentially be extracted compared with traditional observational studies. In EHR-based observational studies, the number of features extracted could easily reach a large number. This may become an issue in the rare-outcomes-common-treatments situation [11] since at least 10 events for every covariate entered into the regression model are suggested [12]. Alternatively, one could use parsimony regression models and only include a subset of covariates in the final model. However, the selection process could be subjective, and researchers may attempt to choose a model that gives a desired or anticipated result since the result can be checked on every run of the covariates selection process [13]. This special characteristic of EHR-based observational studies makes the PSM analysis a good choice besides regression-based methods.

EHR-based observational studies are playing an important role in fighting the current Coronavirus Disease (COVID-19) pandemic. Typically, an effect of an off-label treatment (treatments that were used beyond their original purpose) is of interest, such as the effect of preventing severe outcomes after coronavirus infection. EHR-based observational studies had shown their power in ruling out less promising treatments immediately after the outbreak of COVID-19 [14] and provided evidence to push promising treatments forward to the stage of an RCT. In consequence, COVID-19 studies which are commonly implemented as EHR-based observational studies could significantly benefit from using PSM analysis. But all potential benefits must be based on the correct use of the PSM technique. Thus, it is crucial to have an overall understanding of the *status quo* of methods used and the reporting quality in the COVID-19 studies using the PSM analysis.

In this study, we systematically reviewed COVID-19-related studies that used PSM analysis as an approach to adjusting confounders. We focused on research that studied an effect of a treatment in a COVID-19-positive population and COVID-19-related severe outcomes as a primary outcome. The primary objective of this study is to provide an overall quality assessment of the methods used for COVID-19-related studies using PSM analysis.

## Method

The process of this study was performed primarily based on the 2020 Preferred Reporting Items for Systematic Reviews and Meta-analysis (PRISMA) statement [15] and was adapted for reviewing methodology in the PSM analysis.

### Search strategy

A search was conducted on 1$^{st}$ June 2021 on PubMed to identify COVID-19-related studies that use the PSM analysis. We searched for articles with "COVID" or "Coronavirus Disease" in the title or abstract and "propensity score" in the title or abstract". The search was limited to publications between 2020 and 2021.

### Selection process

A study was included if both two following inclusion criteria were satisfied: 1. The purpose of the study was to assess the therapeutic or harmful effect of a treatment/intervention/exposure in a COVID-19 positive population and use one of the following COVID-19-severity-related outcomes as a primary outcome: hospitalization, ventilation, ICU admission, and death, or a severity index calculated based on those four outcomes. 2. The propensity score matching was used as one of the methods to adjust possible confounders (i.e., studies that only used other propensity score methods rather than matching were excluded). Studies were excluded if they were reviews/meta-analyses, methodological studies, research letters/editorials, or conference abstracts. Title and abstracts were screened to assess the eligibility of the first inclusion criterion. One reviewer (C. G.) independently completed the whole selection process, and no automation tools were used.

## Data extraction and measurement definition

Eligible studies were checked manually to extract information that can be classified into three major categories: article publication information, characteristics of study design, and components of PSM analysis. Article publication information included: title of the article, authors, year/month of publication, journal of publication. The following characteristics of study design are included: study type, target population, intervention, control group, primary outcomes. PSM analysis components included the reporting of covariates in the PS model, justification of covariates selection, handling of missing data, PS model, matching algorithm, matching ratio, caliper specification, replacement option, balance diagnostics, use of paired statistical methods, reporting of estimated treatment effect type, and propensity score matching software.

The covariates were extracted from the "Methods" part first, and if no such information was provided in the "Methods" part, covariates were extracted from tables and supplementary materials that compared baseline characteristics between the post-matched treatment and control groups [7]. The reporting of covariate was denoted as "Yes" if covariates can be easily found and provided with details in the "Methods" part or can be extracted from other text, tables, and supplementary materials, and as "No" if no detail about covariates can be found anywhere in the article.

The justification of covariates selection was denoted as yes if the author reported details about any of the two covariate selection approaches: 1). based on previous studies or guidelines 2). covariates were selected based on variables that were potentially associated with treatment, outcome, or both [9]. Selecting available variables that were significantly different between two groups based on some statistical tests as covariates in the PS model will be treated as "No" [16] because neither it provide information that allows readers to independently evaluate the completeness of confounders nor such kind selection is recommended [17] [18] [19] in PSM analysis.

The presence of statistician co-author was defined as "Yes" if any author had an affiliation with a department of biostatistics, epidemiology, or mathematics [16, 20].

Paired or robust statistical methods were used for post-matched data if 1) the author explicitly mentioned that a "paired", or "matched", "robust" statistical test was used or 2) one of the following methods was used: paired t-test, Wilcoxon signed-rank test, McNemar's test, Cox proportional hazards models stratified on the matched paired, methods with robust standard errors account for clustering in matched pairs (e.g. Cox proportional hazards with robust standard errors), conditional methods that account for matched-pairs design (e.g. conditional logistic regression, generalized estimating equation methods), and bootstrapping [19, 21-24].

Details about definitions and examples of all extracted characteristics of study design and components of PSM analysis were listed in Table 1 [16].

Table 1 Definitions and examples of characteristics of study design and components of PSM analysis

| Components | Definition | Example |
|---|---|---|
| **Study design** | | |
| Study type | The study type was extracted from included studies and was inferred by the reviewer if not provided explicitly in the original article (e.g., prospective/retrospective cohort study or cohort study if could not decide prospective or retrospective, case-control study, cross-sectional study, and clinical trials). | "Retrospective cohort study in 2121 consecutive adults with acute inpatient hospital admission between 4 March and 29 August 2020 with confirmed or suspected COVID-19 in a large academic health system" [25] |
| Sample size | The number of COVID-19 positive subjects in a study before any handling of missing data. If no such information was provided, then the largest number of COVID-19 positive subjects was used. This measurement only roughly reflected the scale of the study. | "A retrospective cohort study on 1014 patients with confirmed COVID-19 diagnosis." [26] |
| Treatment | The effect of which is of interest. | "Patients were identified as treated with hydroxychloroquine if they received at least one dose within 48 hours of admission. The control group for this analysis consisted of patients who were not treated with hydroxychloroquine within 48 hours of admission." [27] |
| Control | The reference group that the effect of treatment is compared with | |
| **PSM analysis components** | | |
| Covariates | Covariates are predictor variables in the PS model. Covariates are not necessarily all confounders. But ideally, covariates should include all true confounders and include as few non-confounders as possible. | "Those COVID-19 patients who were diagnosed with asthma (N = 21) in the cohort before our study were enrolled and 100 patients without asthma were matched by propensity score at an approximate ratio of 1:5 based on age and sex and comorbidities including hypertension, diabetes, coronary heart disease et al (Table 1). "[28] |
| Reporting of covariates | The reporting of covariates was denoted as "Yes" if covariates can be easily found and provided with details in the "Methods" part or can be extracted from other text, tables, and supplementary materials, and as "No" if no detail about covariates can be found anywhere in the article. This variable represents the quality of reporting covariates that were used in the PSM analysis. | |
| Justification of covariates selection | Whether how covariates are selected is mentioned. A well-founded covariates selection process could prevent omitting true confounders. It also allows readers to independently evaluate the completeness of confounders included in the PS model. | "The criteria to include variables in this model were based on those potentially affecting the likelihood of outcome occurrence and receipt of study treatments and were performed based on subject matter knowledge with the help of a direct acyclic graph (DAG) (e-Fig. 1 in Additional file 1) [35, 36]" [29] |
| Handling of missing data | How missing data is handled before being used to calculate propensity scores. Possible methods include pairwise deletion/ complete data analysis (only use observations with no missing data), variable dropping (dropping variables with too many missing values), multiple imputation, etc. | "We used nonparametric missing value imputation, based on the missForest procedure in the R, to account for the missing data on the laboratory variables of increased CRP, LDL-c, eGFR, ALT, CK, BUN, D-dimer and cholesterol as well as decreased |
| | | lymphocyte counts (Waljee et al., 2013)." [30] |
| PS model | The model that uses the assignment of treatment as the outcome and covariates as predictors to estimate the propensity score. | "To derive propensity scores, we constructed a logistic regression model to predict immunosuppression status by including all patient demographic and clinical characteristics listed in the Covariates section above." [25] |
| Matching algorithm | The method that matches subjects in the treatment group with subjects in control groups. Commonly used methods included the nearest-neighbor matching algorithm (a.k.a. greedy algorithm) and optimal algorithm. | "Propensity score matching was performed to obtain a 1:1 matched cohort using the 'nearest-neighbor' approach without replacement, with a match tolerance of 0.1." [26] |
| Matching ratio | The ratio of subjects in the treatment group that are matched to subjects in the control group (e.g., 1:1, 1:2, 1:3). For some matching algorithms such as full matching, the rate may not be fixed. | |
| Replacement option | If matching is done with replacement, a subject in the control group is allowed to be matched with more than one subject in the treatment group. Otherwise, a subject in the control group is only allowed to be matched to one subject in the treatment group if it is used in matching. | |
| Caliper specifications | The maximum distance (typically the difference of propensity score between subjects is used) is allowed between matched subjects. A commonly used caliper is 0.2 of the standard deviation of logit transformation of propensity score [31]. | |
| Balance diagnostics | The distributions of covariates in the post-matched treatment and control group are supposed to be similar. Conventional statistics tests based on p-value should be avoided and the use of standardized mean differences (SMDs) is recommended [19]. If there is still a noticeable difference remained, one should go back to change the PS model (e.g., adding new covariates, considering interactions between covariates, changing the model, etc.) and repeat this process until a good balance is achieved [18]. A doubly-robust model that includes those covariates that still cannot achieve balance may help. | "The balance of covariates was evaluated by estimating SDs before and after matching, and small absolute value <0.1" [32] |
| Use of paired or robust statistical methods | The analysis methods used for the post-matched sample should account the for paired nature of the sample and the use of an unmatched test tends to be more conservative (fails to reject null hypothesis while it is false) [33]. | "The McNemar test for matched pairs was used to assess whether the risk of in-hospital mortality differed according to the status of AF and relevant subgroups during the patients' hospital stay. 9 Risk ratios for the matched-pair analyses were computed and corresponding 95% CIs were estimated using |
| | | 1000 bootstrap resamples with replacement." [34] |
| Reporting of estimated treatment effect type | There are two commonly used types of treatment effects: the average treatment effect (ATE) and the average treatment effect in treated (ATT).<br><br>ATT is the average effect of treatment for those who ultimately receive the treatment and ATE is the average effect of treatment for a sample that is randomly selected from a population. ATT and ATE are usually different in an observational study for people who receive treatment are generally different than the general population. | "The average treatment effect on the treated was calculated with robust Abaide-Imbens standard errors." [35] |
| Propensity score matching software | The software used to do the propensity score matching. | "We performed the propensity score matching using the R software, with the MatchIt and Cobalt libraries." [36] |

### Data analysis

A descriptive table was created to describe distributions of variables from article publication information, study design characteristics, and PSM analysis components, overall and by the presence of statistician co-author. Categorical variables were described with frequencies and percentages. Fisher's exact test was used for testing differential distributions of variables between studies with and without statistician co-author.

All reported P-values were two-sided and a test result with a p-value of less than 0.05 was considered as statistically significant. Since this was an exploratory study, multiple comparisons were not adjusted. All analyses were conducted using R software version 4.0.0 (R Development Core Team, Vienna, Austria).

### Results

A total of 385 citations was retrieved from PubMed and 29 were excluded during the screening: 12 for being duplicates, 14 for their publication types were in exclusion criterion, 4 for no digital access from the Texas Medical Center Library System. For the 356 articles that were retrieved, 184 were further excluded: one for language is not English, one for no access to the supplementary materials, 127 for not meeting the first inclusion criterion (study topics requirement), and 55 for not meeting the second inclusion criterion (use the propensity score but not the Propensity Score Matching) (Fig 1). Finally, a total of 171 articles were included.

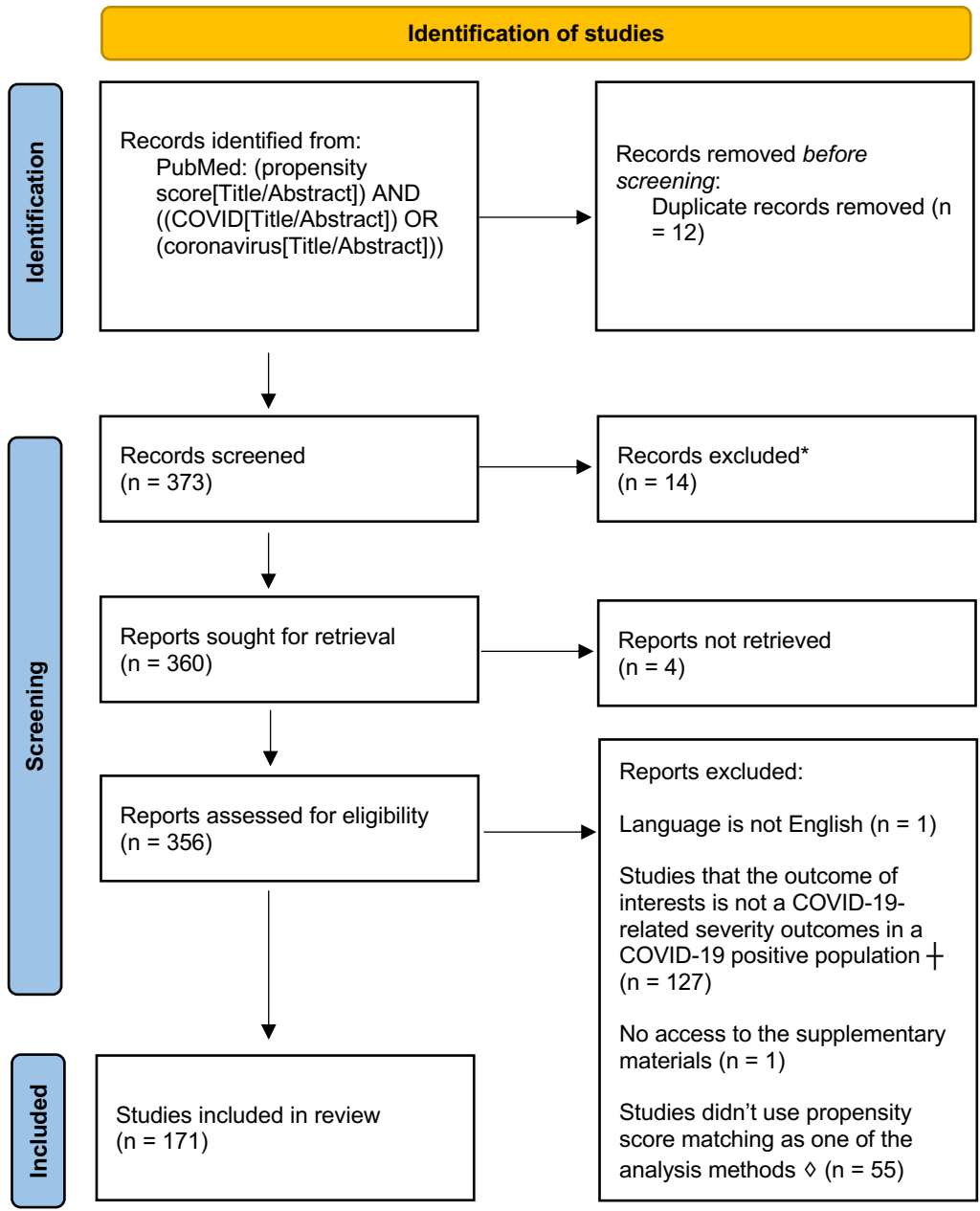

Fig 1. PRISMA flow diagram of the searching strategy and selection process

*\* Studies that were methodology research, review or (and) meta-analysis, research letter/editorial, and conference abstract*

*╂Studies that didn't meet the first inclusion criterion: The purpose of the study was to assess the therapeutic or harmful effect of a treatment/intervention/exposure in a COVID-19 positive population and use one of the following*

*COVID-19-severity-related outcomes as a primary outcome: hospitalization, ventilation, ICU admission, and death, or a severity index calculated based on those four outcomes.*
*◊ Studies that didn't meet the second inclusion criterion: Propensity score matching was used as one of the methods to adjust possible confounders*

### Study characteristics
One-hundred-and-fifty studies (87.72%) were cohort studies (mostly retrospective), thirteen (7.60%) were case-control studies, and the rest of the studies (4.68%, n=8) were cross-sectional, or clinical trials or study type cannot be inferred. The sample size of included studies had a wide range, but most studies had a sample size of COVID-19 positive subjects below 10, 000 (85.38%, n=146), only a very small proportion of studies had a sample size larger than 100, 000 (1.75%, n=3).

### Quality of PSM analysis methodological reporting
Only fifty-four (31.58%) studies explicitly reported how missing data was handled, and the multiple imputation method was the most popular method among them. Although almost all studies (95.91%, n=164) reported what covariates were used in the PS model, only forty-five studies (26.32%) provided a reasonable justification for covariates selection. More than half (60.23%, n=103) studies didn't explicitly provide the model they used to calculate the propensity score, and non-parsimonious multiple logistic regression was the most dominant (62 out of 68) PS model. Similarly, sixty-nine (40.35%) studies didn't report the matching algorithm and nearest-neighbor matching was the most dominant matching algorithm (95 out of 102) upon reported ones. Around half (57.31%, n=98) studies didn't explicitly report what method(s) they used to check post-matching covariates balance, and among those who provided it, most of them (62 out of 73) used the standardized mean differences (SMDs) as one of the balance diagnosis methods.

### Software and tools used in PSM analysis
R software using MatchIt package was the most popular PS matching software (28.65%, n=49), followed by SPSS (16.86%, n=29), SAS (11.63%, n=20), STATA (9.30%, n=16), and others (6.40%, n=11). The PS matching softwares in the "Others" category included: TriNetX online research platform (n=7), python (n=2), EmpowerStats (n=1), and ATLAS interactive analysis platform (n=1) (Table 2).

### Association between statistician co-author and reporting quality of PSM analysis
Forty studies (23.39%) were considered as having statistician co-author. Studies that had statistician co-author were more likely to report methods used in handling missing data (p=0.037). And it was worth noting that all the case-control studies (n=13) were in the "without statistician co-author" category (p=0.006) and all studies that reported the type of estimated treatment effect (n=6) were in the "with statistician co-author" category (p<0.001).

Table 2. Characteristics of included studies

| Variable | levels | Overall | Statistician co-author | | P-value † |
| --- | --- | --- | --- | --- | --- |
| | | | FALSE | TRUE | |
| Year of publication | | | | | |
| | 2020 | 66 (38.60%) | 49 (37.40%) | 17 (42.50%) | 0.562 |
| | 2021 | 105 (61.40%) | 82 (62.60%) | 23 (57.50%) | |
| Study design type | | | | | |
| | Cannot decided | 4 (2.34%) | 4 (3.05%) | 0 (0%) | |
| | Case-control study | 13 (7.60%) | 13 (9.92%) | 0 (0%) | |
| | Cohort study | 14 (8.19%) | 6 (4.58%) | 8 (20.00%) | 0.006** |
| | Others | 4 (2.34%) | 2 (1.53%) | 2 (5.00%) | |
| | Prospective cohort study | 8 (4.68%) | 7 (5.34%) | 1 (2.50%) | |
| | Retrospective cohort study | 128 (74.85%) | 99 (75.57%) | 29 (72.50%) | |
| Sample size | | | | | |
| | 1 ~ 1000 | 80 (46.78%) | 68 (51.91%) | 12 (30.00%) | |
| | 1001 ~ 10, 000 | 66 (38.60%) | 43 (32.82%) | 23 (57.50%) | 0.035* |
| | 10, 001 ~ 100, 000 | 22 (12.87%) | 17 (12.98%) | 5 (12.50%) | |
| | > 100,000 | 3 (1.75%) | 3 (2.29%) | 0 (0%) | |
| Reporting of handling of missing | | | | | |
| | Yes | 54 (31.58%) | 36 (27.48%) | 18 (45.00%) | 0.037* |
| | No | 117 (68.42%) | 95 (72.52%) | 22 (55.00%) | |
| Reporting of covariates | | | | | |
| | Yes | 164 (95.91%) | 125 (95.42%) | 39 (97.50%) | 1.000 |
| | No | 7 (4.09%) | 6 (4.58%) | 1 (2.50%) | |
| Covariate selection justification | | | | | |
| | Yes | 45 (26.32%) | 34 (25.95%) | 11 (27.50%) | 0.846 |
| | No | 126 (73.68%) | 97 (74.05%) | 29 (72.50%) | |
| Propensity score model | | | | | |
| | Logistic regression | 62 (36.26%) | 48 (36.64%) | 14 (35.00%) | |
| | Multiple probit regression | 1 (0.58%) | 0 (0%) | 1 (2.50%) | |
| | Regularized logistic regression | 3 (1.75%) | 2 (1.53%) | 1 (2.50%) | 0.292 |
| | Time-dependent Cox regression model | 2 (1.17%) | 1 (0.76%) | 1 (2.50%) | |
| | Not available | 103 (60.23%) | 80 (61.07%) | 23 (57.50%) | |
| Matching algorithm | | | | | |
| | NNM ‡ | 95 (55.56%) | 76 (58.02%) | 19 (47.50%) | |
| | NNM combined with exact match | 1 (0.58%) | 1 (0.76%) | 0 (0%) | 0.529 |
| | optimal matching | 6 (3.51%) | 5 (3.82%) | 1 (2.50%) | |
| | Not available | 69 (40.35%) | 49 (37.40%) | 20 (50.00%) | |
| Matching ratio | | | | | |
| | 1:1 | 84 (49.12%) | 71 (54.20%) | 13 (32.50%) | 0.061 |
| | 1:2 | 16 (9.36%) | 12 (9.16%) | 4 (10.00%) | |
| | 1:3 | 10 (5.85%) | 8 (6.11%) | 2 (5.00%) | |
| | 4 or more control matched per treatment unit | 11 (6.43%) | 9 (6.87%) | 2 (5.00%) | |
| | multiple ratios used | 3 (1.75%) | 1 (0.76%) | 2 (5.00%) | |
| | Not available | 45 (26.32%) | 29 (22.14%) | 16 (40.00%) | |
| | Unclear | 2 (1.17%) | 1 (0.76%) | 1 (2.50%) | |
| Replacement option | | | | | |
| | With replacement | 4 (2.34%) | 4 (3.05%) | 0 (0%) | |
| | Without replacement | 34 (19.88%) | 26 (19.85%) | 8 (20.00%) | 0.786 |
| | Not available | 133 (77.78%) | 101 (77.10%) | 32 (80.00%) | |
| Use SMD as one of the balance diagnosis methods | | | | | |
| | Yes | 62 (36.26%) | 49 (37.40%) | 13 (32.50%) | |
| | No | 11 (6.43%) | 9 (6.87%) | 2 (5.00%) | 0.768 |
| | Not available | 98 (57.31%) | 73 (55.73%) | 25 (62.50%) | |
| Use paired or robust method | | | | | |
| | Yes | 17 (9.94%) | 11 (8.40%) | 6 (15.00%) | 0.234 |
| | No | 154 (90.06%) | 120 (91.60%) | 34 (85.00%) | |
| Reporting of estimated treatment effect type | | | | | |
| | Yes | 6 (3.51%) | 0 (0%) | 6 (15.00%) | < 0.001*** |
| | No | 165 (96.49%) | 131 (100.00%) | 34 (85.00%) | |
| PS matching software | | | | | |
| | R software | 49 (28.65%) | 36 (27.48%) | 13 (32.50%) | |
| | SPSS | 29 (16.96%) | 26 (19.85%) | 3 (7.50%) | |
| | SAS | 20 (11.70%) | 14 (10.69%) | 6 (15.00%) | 0.249 |
| | STATA | 16 (9.36%) | 10 (7.63%) | 6 (15.00%) | |
| | Others § | 11 (6.43%) | 10 (7.63%) | 1 (2.50%) | |
| | Not available | 46 (28.65%) | 35 (26.72%) | 11 (27.50%) | |

*† P-values were calculated by Fisher's Exact Test*
*‡ NMM: Nearest-neighbor matching*
*§ The propensity score matching software in the "Others" category included: TriNetX online research platform (n=7), python (n=2), EmpowerStats (n=1), and ATLAS interactive analysis platform (n=1).*

# Discussion

This is the first systematic review of the PSM analysis in COVID-19-related studies. Although we have provided several important components that should be reported and recommended reporting format in Table 1, it by no means served as a complete list for reporting PSM analysis. Readers can check Yao, X.I. (2017) [7] and Thoemmes, F.J. (2011) [37] for a more detailed list for reporting PSM analysis, and MatchIt manual and Staffa, S.J. (2018) [38][39] for a step-by-step instruction about how to use propensity score from the beginning. In the rest of this section, we focused on those caveats that were not being given enough attention as they should be.

### PSM analysis in case-control study

It has been shown that the PSM method used in the case-control study was rare in this field. Despite the rarity of such kind of studies, strong evidence was found about the association between using PSM in the case-control study and biostatistician co-author in this study. Only a few methodological studies and reviews had discussed using the propensity score in the case-control study [7][40][41].

Two different strategies of using the propensity score were found in those case-control studies. Some studies modeled the relationship between study outcome (instead of exposure) and covariates to calculate propensity score. Other studies calculate propensity scores using models that used exposure as the dependent variable like in a cohort study. There are some caveats no matter which approach is used. For the first strategy, propensity score matching was proposed as a method to achieve the "study-base" principle (cases and controls should come from the same population) [42] [43]. Allen, A.S. and G.A. Satten (2011) [44] tried to give a mathematical proof of such application but was criticized by Greenland, S. (2018) [45] for "trohoc fallacy". For the second strategy, the propensity score cannot be calculated directly using the entire sample, and the appropriate methods of calculating the propensity score are much more complicated. Although those methods have been proven for good large-sample properties, artifactual effect modification and reduced ability to control for confounding factors are concerned when the sample size is not large enough [41]. The propensity score methods are less straightforward to use in case-control studies and should be avoided unless there is a good reason for it.

### Reporting of estimated treatment effect type

Another key aspect that was rarely reported (3.51%, n=6) but showed a strong association with the presence of biostatistician co-author was the reporting of estimated treatment effect type. The Average Treatment in Treated (ATT) often confuses with the Average Treatment Effect (ATE), which are the two most used types of estimated treatment effect (a.k.a. estimands, a quantity that is to be estimated in a statistical analysis). The ATT is the average effect of treatment for those who ultimately receive the treatment. The ATE is the average effect of treatment for a sample that is randomly selected from a general population. The ATT and ATE are usually different in observational studies for people who receive treatment are generally different than the general population.

Which estimand to use depends on the target population, which is the population that researchers want to generalize the estimated treatment effect to. While PSM analysis is versatile in estimating the ATT and ATE, usually only the ATT or only the ATE can be estimated when a specific matching method is decided. It should be noticed that the nearest neighbor matching (NNM) method, as the most dominant matching method, is almost always used for estimating the ATT rather than the ATE [46]. Given that most studies in this research field want to generalize the effect to a more general COVID-19 population, the preference of using the NNM method could be a problem. Therefore, it is recommended to clarify whether the ATT or ATE is estimated by the PSM analysis to not only show that the researchers are fully aware of what estimand they aim at and whether correct matching methods are used, but also facilitate appropriate interpretation of the estimated effect.

The two estimands discussed above are restricted to cases without dropping treated units due to the use of caliper or common support restriction. When treated units are dropped, the estimated effect is in an artificial population. Such an estimand is called the Treatment Effect in the Remaining Matched Sample (ATM) other than the ATE or ATT [47]. The use of caliper or common support restriction could improve the quality of evidence for a causal relationship by having a better post-matched covariates balance, but it also reduces the generalizability to a broader population meanwhile. So, there is a trade-off in matching for samples in which good post-matched covariates balance cannot be achieved when all the treated units are kept [7]. Although there is a recommendation for using 0.2 of the standard deviation of the logit of the propensity score as the caliper [31], there is no need for using a caliper if a decent post-matching balance can be achieved directly (SMDs < 0.1 is commonly used [39, 48, 49]). And if a caliper is used, reporting the sample sizes of both the treated cohort and control cohort before and after matching can help readers understand the extent of the trade-off that is made [50].

### Paired or robust methods
Like similar studies in other fields, we also observed that only a small proportion (around 10%) of studies used paired methods or robust methods for post-matching analysis in this field [16] [23]. Using paired analyses such as paired t-test was advocated by Austin (2008) for resolving the dependence in post-matched data in the early days [19]. The benefit of using a paired analysis was shown in many simulation studies [51][52]. Alternative paired analyses for commonly used unmatched analyses include using paired-t-test for two-sample t-test, Wilcoxon signed-rank test for Wilcoxon rank-sum test, McNemar's test for Pearson's chi-square test, Mantel-Haenszel test for Fisher's exact test, Stratified log-rank test with stratification on matched pairs for log-rank test, and Cox proportional hazards model with stratification on matched pairs for Cox proportional hazards model [20, 47, 53]. Robust variance estimation methods can be applied when covariate adjustment was used for the remaining imbalance in covariates after matching, such as generalized estimating equation (GEE) methods with variance-structure accounting for matched-pairs design [18]. Robust variance estimation methods are also flexible in accommodating non-fixed ratio matching, matching with replacement, and marginal or conditional effect estimation. Readers can check Greifer, N (2021) for more information about which method to use for a given PSM setting [47].

### Limitation of this systematic review
There were a few limitations in this study. First, the inclusion criterion doesn't cover all COVID-19-related studies since we restricted the included studies to a specific type. However, this type of study should account for most COVID-19-related studies. Second, our literature search was based only on PubMed database. This could moderately impact the representativeness of the literature included in this study [54].

Code and the complete list of reviewed papers:
https://github.com/FDUguchunhui/covid_research_propensity_score_review